# Librational motion of asymmetric rolling bodies and the role of friction force

P. Onorato, M. Malgieri, P. Mascheretti and A. De Ambrosis

Department of Physics  University of Pavia, Via Bassi 6, I-27100 Pavia, Italy

pasquale.onorato@unipv.it

We designed a very simple "asymmetric rolling spool" and we studied it experimentally, in the librational regime, by using a video camera and video analysis software. Comparison of experimental results and theoretical analysis allows addressing critical aspects related both to the characteristic of the oscillatory anharmonic motion of the system and to the role of friction forces in determining it.

For that reasons the "asymmetric rolling spool" can be presented to undergraduate students as an interesting 'case study' where to apply their physics knowledge to understand the behavior of real objects.

## I INTRODUCTION

The pure rolling motion of a homogeneous cylinder is a typical example dealt with in most mechanics textbooks, but only few of them address the same problem when the center of mass of the cylinder lies out of its axis.

Some recent papers discussed the dynamics of a not azimuthally symmetric body rolling on a horizontal[1,2,3] or inclined plane[4,5] showing how an in depth study of asymmetric rolling bodies can stimulate a more accurate comprehension of the physics of rolling.

The purpose of this paper is to discuss the librational motion of a rolling spool, where an attached extra mass breaks the azimuthal symmetry. In such system, during each oscillation, the friction force switches sign and changes in magnitude, producing the non-uniform acceleration of the center of mass. Observing and analyzing these oscillations can help students overcome the idea that friction always has a resistive effect, generating a force that invariably opposes motion. Moreover a careful quantitative discussion can show (see section V) that the normal force, which determines the value of the friction force, is not equal to the weight of the spool and depends on its motion. As a consequence, the asymmetric rolling spool (ARS) can lose grip and start slipping (for the same reason an ARS can jump when it is rolling down an inclined plane.[5]

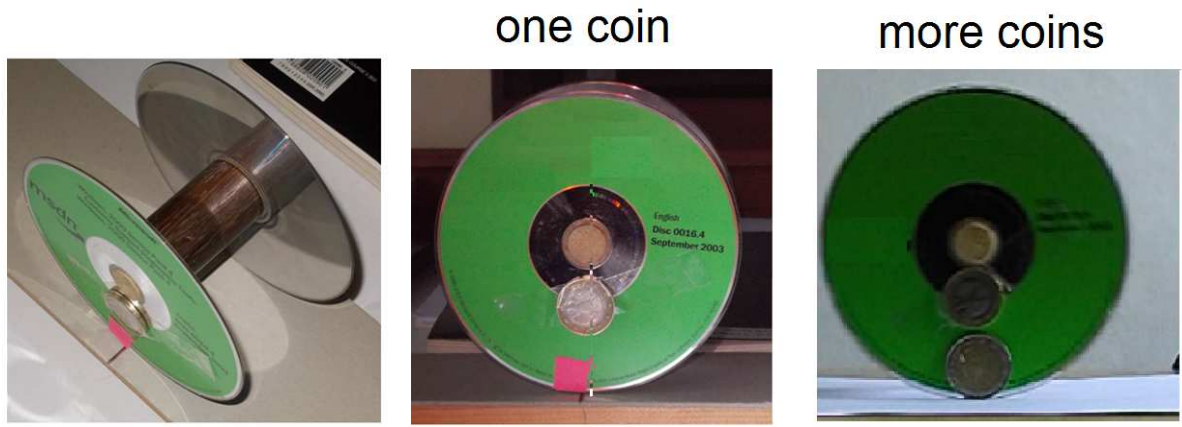

FIG. 1 The rolling spool (made with two CDs and a wooden cylinder) with extra masses (coins) breaking the symmetry of the system.

**II THE ASYMMETRIC ROLLING SPOOL** .

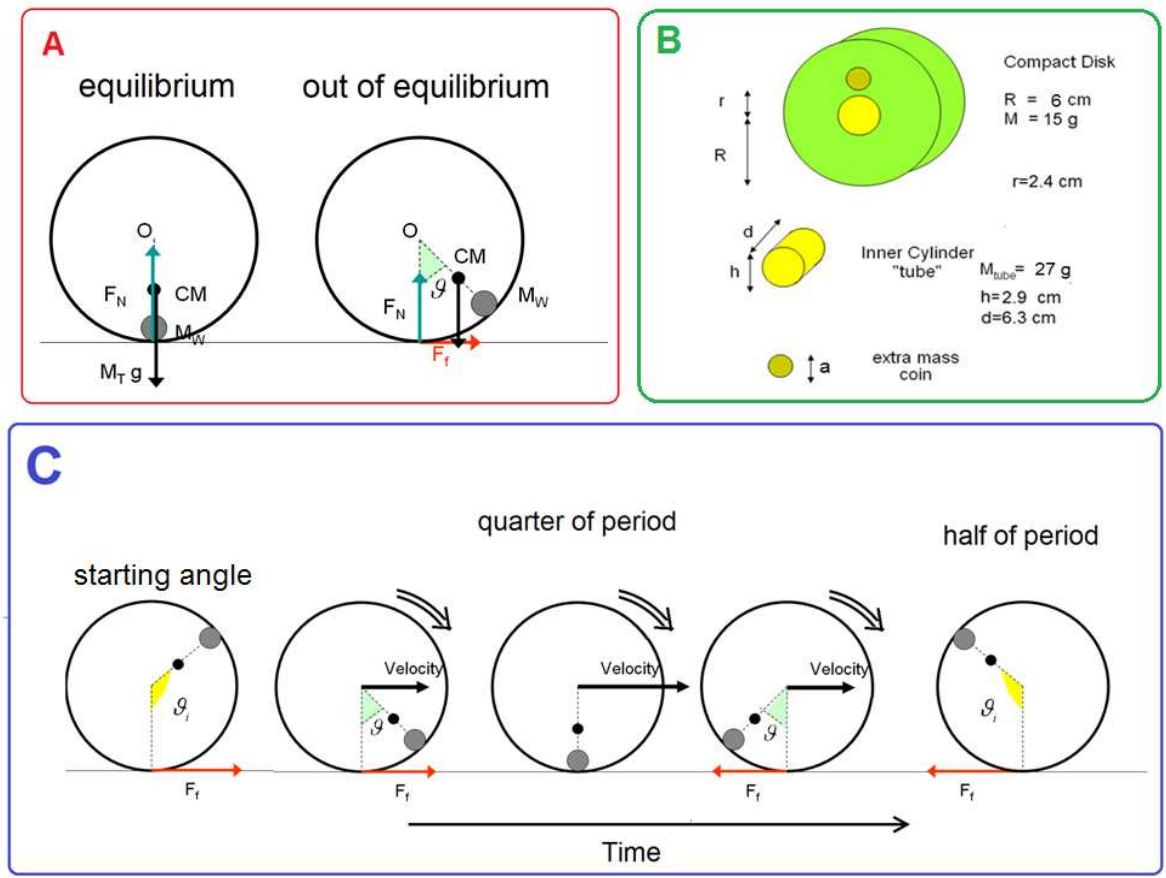

FIG. 2 (A) Forces on the rolling spool: gravity on the center of mass, normal force, and friction force, the only one acting along the horizontal plane. (B) Schematic representation of the spool displaying the values of its geometrical parameters. (C) During each oscillation of the rolling spool the friction force changes accelerating the system.

The mechanical system (Figure 1), consists of a symmetric body of mass $M_0$, and radius R with some extra masses $M_i$ located at distance $r_i$ from the center. From a theoretical point of view, the only simplifying assumption we make is to ignore possible vibrations and oscillations of the spool due to the displacement of the center of mass of the system in the "depth" dimension (referring to figure 2). Experimentally, this effect only becomes significant when the pure rolling condition starts to fail; and even then, it can be countered by placing the additional weights on both sides of the spool.

In the experiments we will analyze devices where we added from 1 to 5 extra masses (coins) located at different distances from the geometric center to study systems with different degrees of asymmetry. In Table 1 the displacements of each added mass from the center are reported, together with the resulting position of the center of mass.

|         | Added Mass $M_i$ (g) | $r_i$ (cm) | $d_{cm}$ (cm) | $\lambda$ |
|---------|----------------------|------------|---------------|-----------|
| 1 coin  | 7.5                  | 2.4        | 0.28          | 0.073     |
| 2 coins | 8.5                  | 4.7        | 0.75          | 0.19      |
| 3 coins | 8.5                  | 4.7        | 1.16          | 0.29      |
| 4 coins | 7.5                  | 4.7        | 1.46          | 0.36      |
| 5 coins | 7.5                  | 4.7        | 1.71          | 0.41      |

**Table 1** Position and mass of each coin, position of the center of mass and asymmetry parameter

All measurements were made by acquiring videos of the motion of the spool through a 30 fps video camera, and then analyzing the videos using the Tracker software[6,7]. We recorded videos of the spool rolling on a horizontal plane and performing free oscillations.

**III A FIRST LOOK ON THE ROLLING MOTION AND THEORETICAL DESCRIPTION**

As it is shown in Figure 1, when the center of mass of the ARS lies on the vertical axis passing through its geometrical center, the system is in a stable equilibrium position. If the spool is placed at rest with a starting angle $\theta_i$ (angle between the vertical axis and the line through the geometrical center and the center of mass) the ARS moves under the action of 3 forces: gravity, the normal reaction force, and friction force. Only the latter force acts along the horizontal plane and is responsible for the horizontal acceleration of the center of mass (i.e. of the whole body).

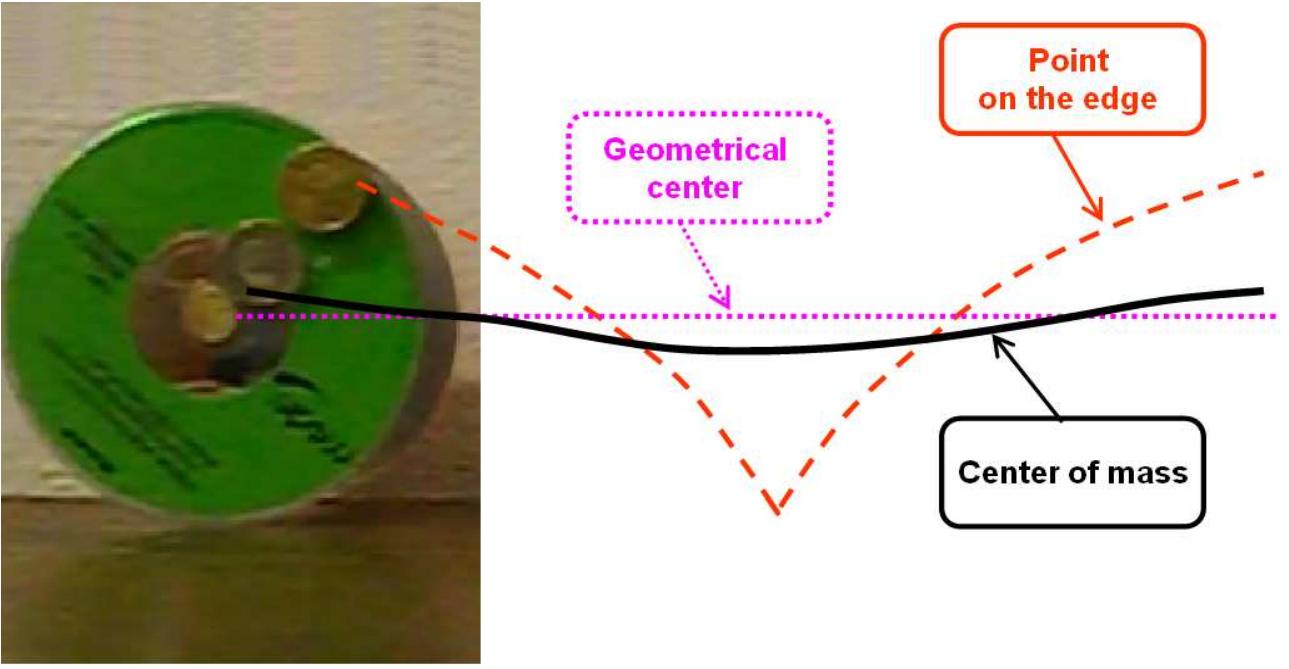

FIG. 3 Trajectories of 3 points on the rolling body acquired with TVA: the point on the edge (cycloydal dashed line) the center of the body and the center of mass (black line).

Thus the ARS oscillates between two extreme positions of maximum displacement, under the action of a time dependent friction force, which changes both in magnitude and direction. Notice that if the friction force is not large enough, the spool has not a pure rolling motion, but it rolls and slips. Using Tracker we can measure the paths followed by different points of the rolling spool. As it is shown in Figure 3, the center of mass oscillates along the vertical direction. Its trajectory (black line) is compared with the cycloidal path of the point on the edge.

The simplest way to obtain the equation of motion of the ARS is through energy conservation. In order to describe the dynamical behavior of the system, we first write the general form of the kinetic energy in the laboratory frame:

$$K(\vartheta,\dot\vartheta) = \frac{1}{2}\left[I_{core} + I_{M_W} + M_{core}R^2 + \sum_i M_i\left(R^2 + r_i^2 - 2Rr_i\cos\vartheta\right)\right]\dot\vartheta^2 = \frac{I_{eff}(\vartheta)}{2}\dot\vartheta^2 \quad (1)$$

Where $M_{core} = M_{CD} + M_{tube}$ in the notations of Figure 2 B.

The center of mass is located at a distance $d_{CM} = \dfrac{\sum_i r_i M_i}{\sum_i M_i + M_{core}} = \dfrac{\sum_i r_i M_i}{M_{TOT}}$ from the center of the cylinder, its coordinates are $x_{cm} = d_{CM}\sin\vartheta - R\vartheta$ and $y_{cm} = -d_{CM}\cos\vartheta$ and, an effective angle-dependent moment of inertia can be identified as

$$I_{eff}(\vartheta) = I_0\left(1 - \lambda\cos\vartheta\right) \quad (2)$$

where we introduced an asymmetry parameter, $\lambda = \dfrac{2\sum_i M_i r_i R}{I_0} = \dfrac{2 M_T R}{I_0} d_{CM}$, which vanishes for a symmetric body. Note that $I_{eff}(\vartheta)$ in eq. (2) corresponds to $I_P(\vartheta)$ as calculated by ref. [5] and [1] for different systems.

As we discussed above, the forces acting on the rolling spool are gravity, normal force and static friction. We can set the equilibrium position, $x_{eq}$ as origin of the x axis, directed along the plane, corresponding to $\vartheta = 0$ so that, when the spool rolls without slipping, $x_c = R\vartheta$ and the velocity of the geometrical center of the disk is $v_C = R\dot\vartheta$. The potential energy of the system is given by

$$U(\vartheta) = M_T g d_{CM}(1 - \cos\vartheta) \tag{3}$$

Combining equations (2) and (3) one obtains[5]

$$\dot\vartheta(\vartheta, \vartheta_i)^2 = 2\frac{(E - U(\vartheta))}{I_{eff}(\vartheta)} = 2\frac{M_T g d_{CM}[\cos(\vartheta) - \cos(\vartheta_i)]}{I_0(1 - \lambda\cos(\vartheta))}, \tag{4}$$

where the second expression is valid if, as in our case, the spool is initially at rest and $\vartheta_i$ is the starting amplitude. To find the equation of motion we multiply Eq. (4) by $I_{eff}(\vartheta)$, we derive both sides with respect to time, and divide the result by the angular velocity $\dot\vartheta$ [1] to obtain

$$\ddot\vartheta = -\frac{\left[g + R\dot\vartheta^2\right]}{I_0(1 - \lambda\cos\vartheta)} M_T d_{CM} \sin(\vartheta) \tag{5}$$

Since *the acceleration of the center of the spool,* $a_x = \ddot\vartheta R$, is not directly proportional to the displacement from equilibrium position, the oscillation period T strongly depends on the amplitude $\vartheta_0$ of the oscillations and can be expressed as

$$T(\vartheta_0) = \oint \frac{d\vartheta}{\dot\vartheta(\vartheta)}$$

where the integral is along the closed orbit and $\dot\vartheta(\vartheta)$ is given by eq.(4). When the amplitude increases, the period raises. The dependence of the period on the oscillation amplitude is a typical signature of anharmonicity, as is the presence of a critical value for the amplitude, approaching which the period tends to diverge.[8-12]

**IV EXPERIMENTAL RESULTS**

In this section we discuss how the degree of asymmetry of the system is a key parameter for determining the small oscillation value of the period, and the form of its amplitude dependence.

## IV a. Anharmonic motion: amplitude dependent periods

The experimental results for the measured period T as a function of the amplitude of oscillation are reported in Figure 4(A) where points correspond to the experimental results, while the dashed line corresponds to the theoretical values (obtained by a numerical integration). A good agreement between predicted and measured values is found and shows the typical features of anharmonic motion.

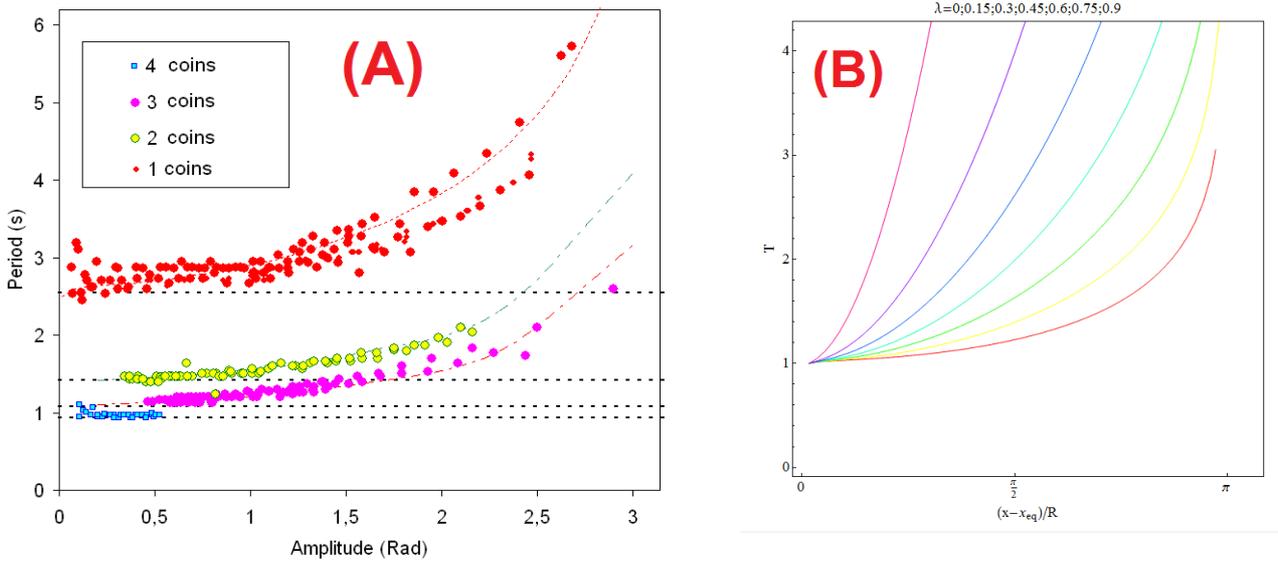

FIG. 4 Experimental results of the measured period $T$ obtained from TVA. Circles correspond to the experimental results while the dashed line corresponds to the theoretical result (black the asymptotic values for small amplitudes). (B) Theoretical values for the ratio of the calculated period $T$ to the asymptotic value $T_0$ as a function of the amplitude of oscillation.

In Fig. 4B we show the numerically computed periods as functions of amplitudes. Anharmonicity is a consequence of a strongly non linear "effective restoring force" i.e. the acceleration of the body is not linearly proportional to the displacement from the equilibrium position. This acceleration, which as we know is due to friction force, can be easily measured thanks to Tracker. In Figure 5 the acceleration of the geometrical center is represented as a function of the linear displacement $x$ from the equilibrium position $x_0$. The graph shows a strongly non linear behavior, fitted by a curve $a_c = A\sin(\frac{x_c}{R_E} + \delta\vartheta)$. The sinusoidal relation of acceleration vs. x can be extrapolated from eq. (5) in the limit of small values of $\lambda$:

$$a_c = R\ddot{\vartheta} \approx \frac{\lambda}{2} g \sin(\vartheta) = \frac{\lambda}{2} g \sin(\frac{x_c}{R}), \qquad (6)$$

This result is in a very good agreement with the measured values plotted in Figure 5.

In table 2 we report the fit parameters obtained with Tracker for acceleration versus position data.

**Acceleration versus position: best-fitting parameters**

|  | Amplitude of acceleration A (m/s²) | $\frac{\lambda}{2}g$ (m/s²) | $\delta\vartheta$ (rad) | $R_E$ (m) |
|---|---|---|---|---|
| 1 coin | 0.35±0.05 | 0.37±0.03 | 0.1±0.1 | 0.059±0.005 |
| 2 coins | 1.1±0.1 | 0.95±0.05 | 0.05±0.1 | 0.057±0.005 |
| 3 coins | 1.6±0.1 | 1.45±0.1 | 0.1±0.1 | 0.056±0.005 |

**Table 2** Fitting parameters for different systems

Results are consistent with the theoretical values calculated starting from the approximate formula in eq. (6). In fact the values of $\delta\vartheta$ are consistent with 0 and the ones of $R_E$ are consistent with the measured radius (6 cm). Moreover, the amplitude A is well approximated by the value $\frac{\lambda}{2}g$.

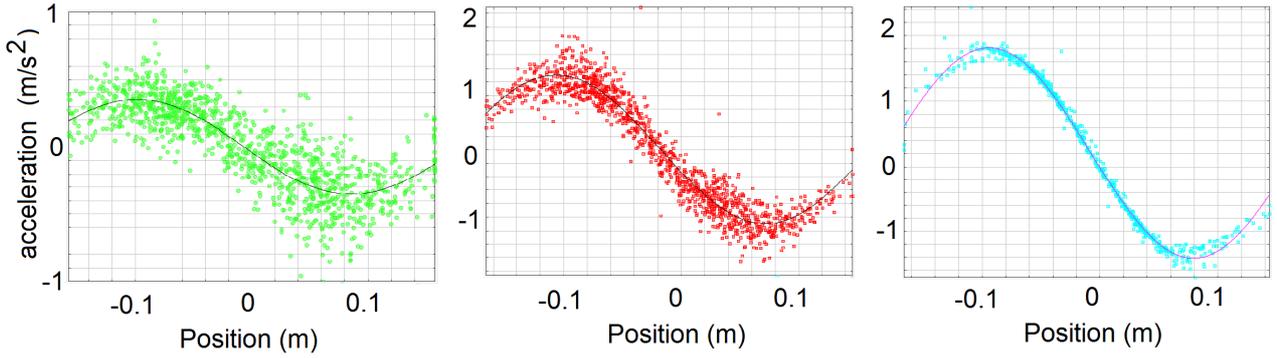

FIG. 5 Measurements and fitting curves for the acceleration of the geometrical center of the spool. From left to right 1, 2, and 3 coins as extra masses.

### IV b. Harmonic limit: small amplitude oscillation periods

For small amplitude oscillations, a "harmonic regime" can be observed with nearly constant period depending on the constructive parameters of the ARS (Fig. 4A).

From a theoretical point of view, in the limit of small oscillation amplitude, we can assume that

$$\ddot{\vartheta} = -\frac{\left[g + R\dot{\vartheta}^2\right]}{I_0(1-\lambda\cos\vartheta)} M_T d_{CM} \sin(\vartheta) \approx -\frac{gM_T d_{CM}}{I_0(1-\lambda)}\vartheta \qquad (7)$$

i.e.

$$\omega = \frac{2\pi}{T} = \sqrt{\frac{g\sum_i M_{Wi} r_i}{I_0(1-\lambda)}g} \qquad (8)$$

Thus the small amplitude (harmonic) oscillations of the asymmetric spool are analogous to those of a simple pendulum with an effective length (radius of gyration[13]) $\ell_{eff} = I_0(1-\lambda) / \sum_i M_{Wi} r_i$.

In Table 3 the theoretical and experimental values of the periods in the harmonic limit are reported showing excellent agreement.

| | $\ell_{eff}$ (m) | $\lambda$ | $T_{exp}$(s) | $T_{th}$(s) |
|---|---|---|---|---|
| 1 coin | 1.5 | 0.073 | 2.5 | 2.5 |
| 2 coins | 0.5 | 0.19 | 1.4 | 1.42 |
| 3 coins | 0.3 | 0.29 | 1.1 | 1.08 |
| 4 coins | 0.2 | 0.36 | 0.95 | 0.93 |

**Periods of small amplitude oscillations**

**Table 3** Periods and effective lengths

## V PURE ROLLING CONDITIONS: A LIMIT TO THE ASYMMETRY OF ROLLING SYSTEMS

As we discussed above and showed in Figure 3, the center of mass also oscillates along the *y* direction, changing the magnitude of the force pressing on the plane (i.e. of the normal force). These oscillations, as discussed in ref. [1], can reduce the friction force below the value needed to ensure pure rolling motion.

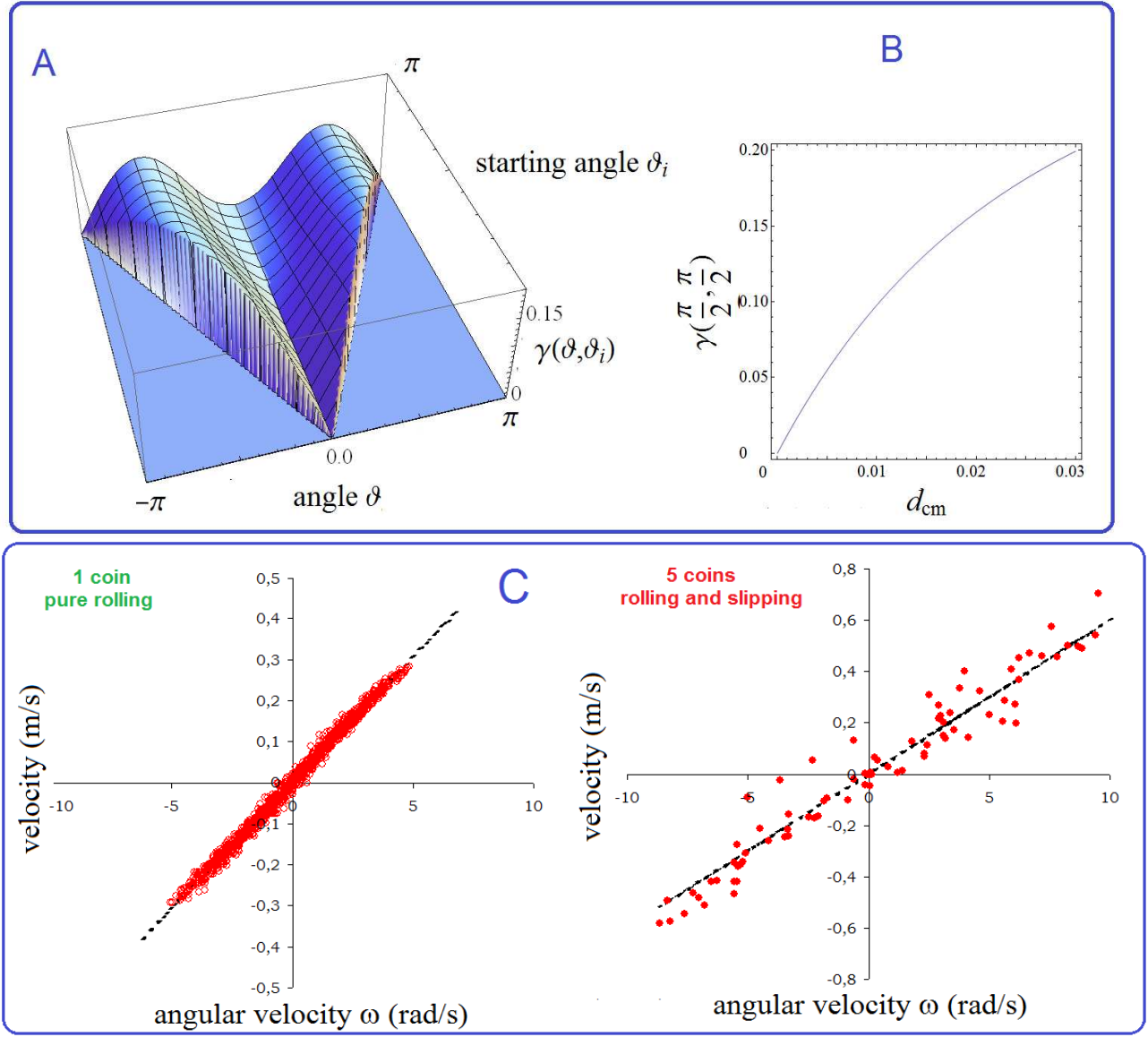

FIG. 6 (A) Values of $\gamma(\vartheta,\vartheta_i)$, for a given value of the others parameters of the system. From the plot, obviously restricted to the region $\vartheta \leq \vartheta_i$ we can deduce that for a starting angle $\vartheta_i \geq \frac{\pi}{2}$, the maximum value of $\gamma(\vartheta,\vartheta_i)$ is reached near $\vartheta \approx \frac{\pi}{2}$, while for $\vartheta_i < \frac{\pi}{2}$ the maximum value of $\gamma(\vartheta,\vartheta_i)$ is reached near $\vartheta \approx \vartheta_i$. (B) $\gamma(\frac{\pi}{2},\frac{\pi}{2})$ as a function of the distance of the center of mass from the geometrical center ($d_{cm}$). Here we suppose that the starting angle is near π/2. (C) The velocity of the geometrical center versus the angular velocity. In the case of pure rolling the ratio $v_{cm}/\omega$ has to be a constant, R, as shown in left panel for 1 coin. This relation is no longer fulfilled for strongly asymmetric rolling bodies as reported in the right panel.

To verify whether the condition of rolling without slipping is satisfied, we analyze the forces acting on the rolling body, which are represented in Figure 2. The friction force, $F_f = M_{tot} a^x_{cm}$, causes the horizontal acceleration of the center of mass, while the normal reaction and gravitational force produce its vertical acceleration, $F_N - M_{tot} g = M_{tot} a^y_{cm}$.

The condition on friction force $|F_f| \leq \mu_s |F_N|$ implies that the rolling spool rolls without slipping if $\mu_s \geq \gamma(\vartheta, \vartheta_i)$, for all values of $\vartheta$ reached during the motion, with

$$\gamma(\vartheta, \vartheta_i) \equiv \frac{|a^x_{cm}|}{|g + a^y_{cm}|} = \left| \frac{-R\ddot{\vartheta} + d_{CM}(\ddot{\vartheta}\cos\vartheta - \dot{\vartheta}^2 \sin\vartheta)}{g + d_{CM}(\ddot{\vartheta}\sin\vartheta + \dot{\vartheta}^2 \cos\vartheta)} \right|,$$

Where $a_x$ and $a_y$ are obtained by differentiating twice, with respect to time, the coordinates $x_c$ and $y_c$ of the center of mass. $\gamma(\vartheta, \vartheta_i)$ depends on the constructive parameters ($R$ and $d_{CM}$) and on the starting angle, $\vartheta_i$. For given $R$ and $d_{CM}$ and for a starting angle $\vartheta_i \geq \frac{\pi}{2}$, the maximum value of $\gamma(\vartheta, \vartheta_i)$ is reached near $\vartheta \approx \frac{\pi}{2}$, while for $\vartheta_i < \frac{\pi}{2}$ the maximum value of $\gamma(\vartheta, \vartheta_i)$ is reached near $\vartheta \approx \vartheta_i$. In Figure 6 we show how $\gamma(\vartheta, \frac{\pi}{2})$ increases with the value of $d_{CM}$ (i.e. as the extra mass becomes larger and / or more displaced from the geometrical center of the disk).

For a static friction coefficient equal 0.14 (as measured for our system with the method described in the appendix) we can estimate that if $d_{cm} > 1.5$ cm the condition $\mu_s \geq \gamma(\vartheta, \frac{\pi}{2})$ is no longer satisfied and the motion is no longer pure rolling.

It is possible to notice the slipping motion of the rolling spool directly by eye inspection; and the measurements with Tracker allow inferring it from the experimental plots of velocity vs. angular velocity (Fig. 5C). For the cases with lower asymmetry (1, 2 or 3 coins) we obtained measurements perfectly placed along a straight line, while for the strongly unbalanced rolling spool (5 coins) experimental measurements are very spread out around the line corresponding the pure rolling condition, suggesting the presence of slipping motion. To confirm this interpretation we propose an alternative approach: to track the trajectory of a point at the edge of the disk. If the spool slips, the cusp of the cycloid drawn by the point on the edge during oscillations is right displaced when the spool is rotating anticlockwise and left displaced when the spool is rotating clockwise. This is the case for a strongly unbalanced spool as we measured by using Tracker. (Fig.7 Left)

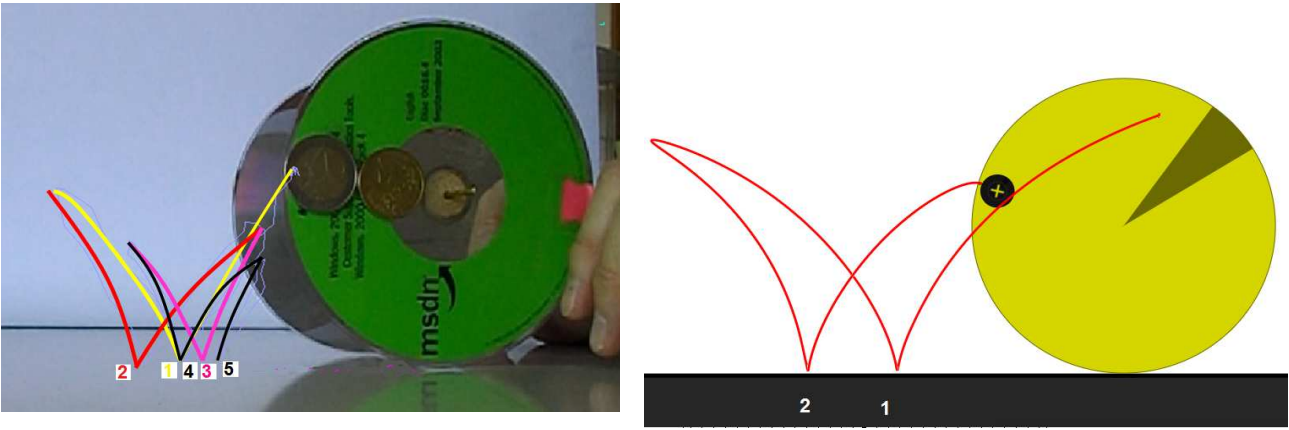

FIG. 7 Evidence of slipping motion from the displacements of the cycloid cusp. Left The experimental data acquired using TVA, and Right the simulation realized using Algodoo software.[14]

For educational purposes it could be useful to simulate the behavior of the sliding and rolling spool by means of a free software like Algodoo, as it is shown in Figure 7, Right.

**VI CONCLUSIONS**

The motion of asymmetric rolling rigid bodies, rolling on a horizontal plane, was discussed in the regime of librational motion. The harmonic and the anharmonic regimes of oscillations where analyzed emphasizing the role of the asymmetry of the body. To compare theoretical predictions with experimental data, we measured the motion of an "asymmetric rolling spool" realized by adding one or more extra masses displaced from the geometric center of the circle. The measured values of the periods agree well with the theoretical predictions.

Moreover the experimental data show that for largely asymmetric bodies the rolling condition is not reliably fulfilled because the intensity of the friction force goes below the needed value to ensure rolling without slipping.

The proposed experiments address relevant topics in the physics curriculum and appear appropriate to be used in high school and undergraduate physics courses.

**APPENDIX: MEASURE OF STATIC FRICTION COEFFICIENT**

To obtain a direct measure of the static sliding friction coefficient we designed the apparatus shown in Figure 8. A pivot passing through the central axis of the spool is fit in a slit in a wooden frame, so that the body cannot start rolling. However the spool can rotate on its axis while slipping on the plane, if a sufficient torque is applied by a horizontal tracking force. Normal force on the spool is varied by means of extra masses hanging below the central axis. Thanks to Tracker we measure the friction force at breakaway $\mathbf{F}_A = -\mathbf{F}_t$ before the spool starts rotating.

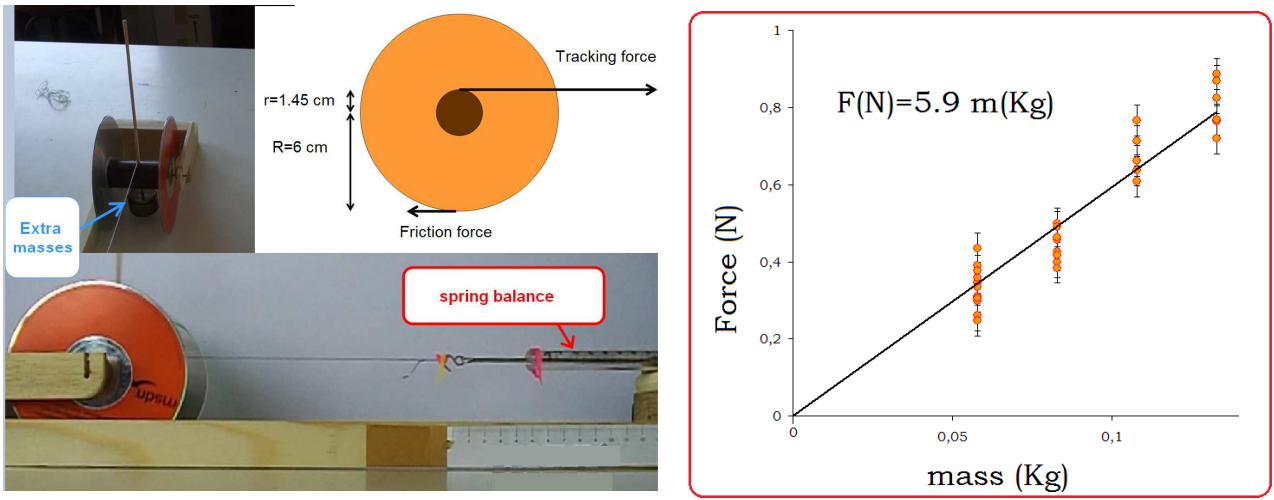

FIG. 8 Left The experimental apparatus for the measure of the friction coefficient and, right, the measurements obtained by changing the extra masses.

From graphs of this force vs. the mass of the system (Fig. 8) we obtain the ratio $F_t/m$ as the slope of the fitting line and evaluate the static sliding friction coefficient from the formula

$$\mu_s = \frac{F_t r}{mgR}$$

where R is the radius of the disk and r the radius of the inner cylinder.

We found $\mu_s \approx 0.14 \pm 0.02$. We can conclude, as observed experimentally, that the 4 coins asymmetric spool does not satisfy the condition of rolling without slipping.